\documentclass[aps,prb,reprint,superscriptaddress]{revtex4-1}
\usepackage[dvips]{graphicx}

\def\rss#1#2{#1_{\mathrm{#2}}}
\def\abs#1{\left| #1 \right|}

\begin{document}

%\preprint{}

\title{Shot noise spectroscopy on a semiconductor quantum dot in the elastic and inelastic cotunneling regimes}


\author{Yuma Okazaki}
\email[]{okazaki.y@lab.ntt.co.jp}
\affiliation{NTT Basic Research Laboratories, NTT Corporation, 3-1 Morinosato-Wakamiya, Atsugi, Kanagawa 243-0198, Japan}
\affiliation{Department of Physics, Tohoku University, Sendai, Miyagi 980-8578, Japan}
\author{Satoshi Sasaki}
\affiliation{NTT Basic Research Laboratories, NTT Corporation, 3-1 Morinosato-Wakamiya, Atsugi, Kanagawa 243-0198, Japan}
\affiliation{Department of Physics, Tohoku University, Sendai, Miyagi 980-8578, Japan}
\author{Koji Muraki}
\affiliation{NTT Basic Research Laboratories, NTT Corporation, 3-1 Morinosato-Wakamiya, Atsugi, Kanagawa 243-0198, Japan}

%\email[]{}
%\thanks{}


\date{\today}

\begin{abstract}
We report shot noise spectroscopy on a semiconductor quantum dot in a cotunneling regime.
The DC conductance measurements show clear signatures of both elastic and inelastic cotunneling transport inside a Coulomb diamond. 
We observed Poissonian shot noise with the Fano factor $F\approx 1$ in the elastic cotunneling regime, and super-Poissonian Fano factor $1<F<3$ in the inelastic cotunneling regime.
The differences in the value of the Fano factor between elastic and inelastic processes reveal the microscopic mechanisms involved in the cotunneling transport.
\end{abstract}

%\keywords{}

\maketitle

%\section{}
%\subsection{}
%\subsubsection{}

 %Introduction of shot noise measurements, the concept of Fano factor is described
Shot noise is a time-dependent current fluctuation reflecting the discreteness of charge carriers.\cite{aBlanterPhysRep00}
Measurements of shot noise can reveal dynamical mechanisms involved in charge transport.
For a conductor in which each successive electron tunneling events can be regarded as noninteracting and uncorrelated, the low-frequency Fourier power spectral density $S$ of time-domain current fluctuation is proportional to the time-averaged current $I$, expressed as $S=2eI$ with $e$ being the charge of the carrier.
This type of noise is called Poissonian shot noise.
For general conductors, the value of $S$ can be either enhanced or suppressed with respect to the case of Poissonian noise due to the presence of electron-electron interactions or correlations.
To characterize the shot-noise properties of various conductors, the Fano factor $F$, defined as $S=2eFI$, is generally used.
$F<1$ ($F>1$) characterizes shot noise suppressed (enhanced) with respect to the Poissonian noise, which is referred to as sub-Poissonian (super-Poissonian) noise.

Transport through quantum dot (QD) structures is influenced by electron-electron interactions and correlations, as manifested by the Coulomb blockade\cite{rKouwenhoven97} and the Kondo effect.\cite{aGoldhaver-GordonNature98, rGrobisCondmat06, rPustilnikJPhys04}
The interactions and correlations underlying these phenomena alter not only the time-averaged net transport but also the shot noise properties,  resulting in both sub-Poissonian and super-Poissonian noise in the Coulomb-blockade and Kondo regimes, respectively.\cite{aKuznetsovPRB98, aSelaPRL06, aGogolinPRL06, aGolubPRB06, aZarchinPRB08, aYamauchiPRL11, aCottetPRB04, aBelzigPRB05, aZhangPRL07, aOnacPRL06, aLossPRL00, aSukhorukovPRB01, aThielmannPRL05, aThielmannPRB05, aWeymannPRB08, aAghassiAPL08}
Electron cotunneling, one of the fundamental transport mechanisms in QD systems, is a higher-order tunneling process which allows a small current to flow in the  Coulomb blockade regime.\cite{aZumbuhlPRL04, aFranceschiPRL01, aSchleserPRL05}
Cotunneling is classified into elastic and inelastic processes;
the former involves only the QD ground state, while the latter accompanies dynamical charge fluctuation between the ground and excited states.
Shot noise measurements in cotunneling regimes enable us to investigate nonequilibrium transport beyond the framework of the linear response theory\cite{aCottetPRB04, aBelzigPRB05, aThielmannPRL05, aThielmannPRB05, aWeymannPRB08, aAghassiAPL08} and to elucidate higher-order two electron correlations, including entanglement and nonlocality.\cite{aLossPRL00, aSukhorukovPRB01}
Previous theoretical investigations have predicted a wide variety of noise behavior in the cotunneling regime; on the other hand, experimental study is still in its infancy.

Onac \textit{et al.}\cite{aOnacPRL06} have reported shot noise measurements on a carbon-nanotube QD.
Exploiting an on-chip noise detector based on photon assisted tunneling in a superconducting junction, they observed super-Poissonian noise in the inelastic cotunneling regime.
However, in the elastic cotunneling regime below the threshold to the inelastic cotunneling, the current level was too low ($<150$ pA) for the shot noise to be resolved by their on-chip noise detection scheme.
For semiconductor QDs, shot noise measurements in the Coulomb blockade regime are made more challenging by the even lower cotunneling current inherent to  semiconductor QDs.\cite{aZumbuhlPRL04, aFranceschiPRL01, aSchleserPRL05}
Gustavsson \textit{et al.}\cite{aGustavssonPRB08} have reported shot noise measurements on a semiconductor QD using a time-resolved charge counting technique, which is capable of resolving the motion of a single electron.
Although they have succeeded in extracting the shot noise of the current in the cotunneling regime, no signal associated with either elastic or inelastic cotunneling was detected, because of the limited time resolution of their measurements.

In this paper, we study the shot noise in a semiconductor QD in the Coulomb blockade regime.
We employ a direct current noise measurement scheme that exploits a cold amplifier and Fourier-transform-based spectral analysis.
In order to facilitate the shot-noise measurement in the Coulomb blockade regime, we fabricated a small QD, in which the level spacing $\Delta E$ can be increased up to 1 meV.
This large $\Delta E$ allows us to apply a large source-drain bias voltage and yield a purely elastic cotunneling current of several 100 pA, which is high enough as compared to the resolution of our noise measurement setup utilizing a cold amplifier.
Simultaneous dc conductance measurements show clear signatures of both elastic and inelastic cotunneling inside a Coulomb diamond.
In the elastic cotunneling regime below the inelastic cotunneling threshold, the measured shot noise shows Poissonian Fano factor $F\approx 1$.
In the inelastic cotunneling regime, in contrast, the shot noise is characterized by a super-Poissonian Fano factor $1<F<3$, indicating that the inelastic mechanism involves subsequent sequential tunneling processes.
Our results, which agree with the theoretically predicted behavior,\cite{aWeymannPRB08} demonstrate that shot noise spectroscopy can elucidate the microscopic mechanisms involved in the tunneling processes that are indistinguishable in dc transport.

%In our system, only two the ground and first-excited states contribute to the enhancement of the Fano factor, as one can find only the lowest excitation state in the region of the $N$-electron diamond [Fig.~\ref{figCotunneling02}(b)].
%This situation provide a good opportunity for quantitative comparisons between the theory and experiment, because the many theories assume that only two single-particle states contribute to the cotunneling.\cite{aThielmannPRL05, aAghassiAPL08}
%In contrast, the previous experiment on a carbon-nanotube QD shows multiple excitation states that contribute to the super-Poissonian shot noise and unsuitable for this comparison.\cite{aOnacPRL06}.
%As a result, the observed Fano factor $F\sim 2.5$ is slightly larger than the theoretical value, which is up to $F=2$.\cite{aThielmannPRL05, aAghassiAPL08}
%We thought that this excess enhancement is probably possible, if the excitation state is strongly coupled to the lead electrodes than the ground state.
%This is because, a large dot-lead coupling of the excitation state increases the  repetition time of the subsequent sequential tunneling, which enhances the Fano factor.
%We note that the previous theory assumes the equivalent coupling strengths for both the ground and excitation states, but no theoretical calculation with a situation for different coupling strengths.

\begin{figure}
 \begin{center}
\includegraphics[scale=0.9]{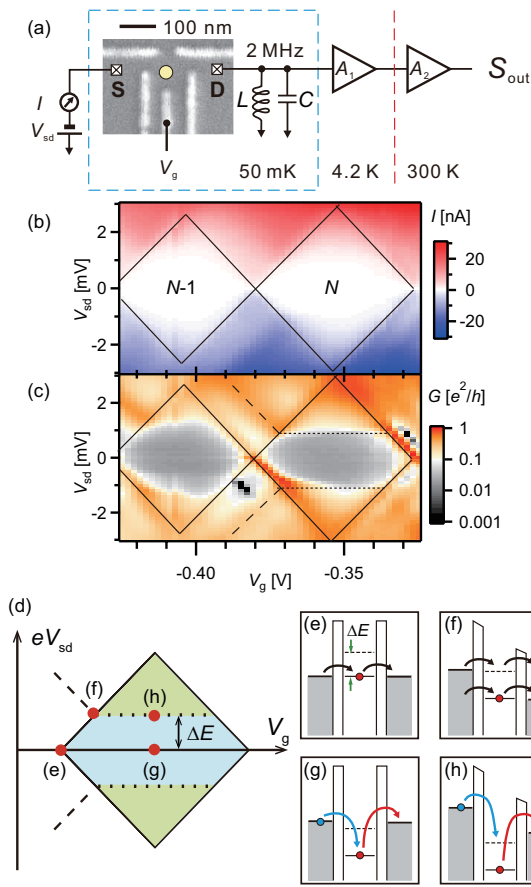}
 \end{center}
 \caption[Setup of inelastic cotunneling and DC spectroscopy]{(Color online). (a) Scanning electron microscope image of the device and the setup for measuring the current noise. (b)-(c): Color plots of (b) current $I$ and (c) differential conductance $G$ of the QD as a function of $\rss{V}{sd}$ and $\rss{V}{g}$. The solid lines depict the edges of the Coulomb diamond associated with the $N-1$ and $N$-electron valleys. The dashed lines outside the diamond show the onset for the sequential tunneling through the excited state. The dotted lines inside the diamond show the threshold voltage for the inelastic cotunneling. (d)-(h): Schematic illustrations of (d) typical Coulomb diamond structure and (e)-(h) tunneling processes relevant in each region shown in (d); (e) sequential tunneling through ground state, (f sequential tunneling through both ground and excited states, (g) elastic cotunneling, and (h) inelastic cotunneling.}
 \label{figCotunneling01}
\end{figure}

Figure \ref{figCotunneling01}(a) shows a scanning electron microscope image of our device and the setup for measuring the current noise.
A single QD is defined in a shallow two-dimensional electron gas (density $n=6.7\times 10^{11}$ $\mathrm{cm^{-2}}$ and mobility $\mu = 4\times 10^{5}$ $\mathrm{cm^2/Vs}$) confined to an $\mathrm{Al_{0.3}Ga_{0.7}As/GaAs/Al_{0.3}Ga_{0.7}As}$ quantum well of 20 nm width, whose center is located 50 nm below the surface.
A QD is formed inside the square area enclosed by the gates, which is $100\times 100$ $\mathrm{nm^2}$ in lithographic dimension.
The transport measurements are carried out in a $\mathrm{^3He}$-$\mathrm{^4He}$ dilution refrigerator with the base temperature $\rss{T}{b}=50$ mK.
The gate voltage $\rss{V}{g}$ is applied to the center gate in order to control the charge states in the QD.
A source-drain bias voltage $\rss{V}{sd}$ is applied to the source contact (S), while the drain contact (D) is shunted to the cold ground through the inductor $L$.
The resulting dc current $I$ is measured at the source side.
Additionally, the differential conductance $G=\mathrm{d}I/\mathrm{d}\rss{V}{sd}$ is simultaneously measured using a standard lock-in technique with 2-$\mathrm{\mu  V}$ ac excitation at 23 Hz.

To measure current noise with higher resolution, we employ a direct current noise measurement scheme that utilizes a cold amplifier and Fourier-transform-based spectral analysis.\cite{ade-PicciottoNature98,  aDicarloRevSciInstl06,  aHashisakaJPhysConfSer08}
In our setup, the current noise from the QD, with its power spectrum density $\rss{S}{QD}$, is fed to an \textit{LC} tank circuit with a center frequency of ${\approx}2$ MHz.
At this frequency, $1/f$ noise is negligible and so only shot noise and thermal noise contribute to the measured $\rss{S}{QD}$.
Our system was calibrated using the thermal noise measured for various resistances at 12 different temperatures from 200 to 700 mK.
The resolution in the current noise is $\delta S \approx 2\times 10^{{-}29}$ $\mathrm{A^2/Hz}$, which corresponds to a Poissonian noise of 60 pA (see also Supplementary Material for more details).

%One or two paragraph for dc conductance spectroscopy
Figure \ref{figCotunneling01}(b) depicts the current through the QD as a function of $\rss{V}{g}$ and $\rss{V}{sd}$. 
The current is suppressed in the two adjacent diamonds indicated by the solid lines.
Inside each diamond, the first-order electron tunneling is energetically prohibited (the Coulomb blockade\cite{rKouwenhoven97}), and the QD has a well-defined number of electrons denoted by $N$ and $N-1$.
Note that $N$ here is even, because the $N{+}1$ valley exhibits the Kondo effect \cite{rGrobisCondmat06} when the dot-lead coupling is slightly increased (data not shown).

Figure \ref{figCotunneling01}(d) illustrates a typical Coulomb diamond structure and the tunneling processes relevant to each region.
Outside the diamond, sequential tunneling allows a current flow through single-particle levels in the QD [Figs.~\ref{figCotunneling01}(e) and (f)].
Inside the diamond, transport through the QD is allowed only via a higher-order tunneling process, the so-called `cotunneling,' in which tunneling of an electron from the source to the blockaded QD occurs simultaneously with tunneling of another electron from the QD to the drain through a virtual state.
Elastic cotunneling conserves the energy of the QD system before and after the process, leaving the QD in the ground state [Fig.~\ref{figCotunneling01}(g)].
In contrast, inelastic cotunneling, which leaves the QD in an excited state [Fig.~\ref{figCotunneling01}(h)], requires an energy supplied from the source-drain bias voltage.\cite{aFranceschiPRL01, aZumbuhlPRL04,aSchleserPRL05}
Accordingly, inelastic cotunneling is allowed only for source-drain biases above a certain threshold $e\rss{V}{sd}$, which is equal to the level spacing $\Delta E$ and is therefore independent of $\rss{V}{g}$ [Fig.~\ref{figCotunneling01}(d)].

\begin{figure}
 \begin{center}
\includegraphics[scale=0.8]{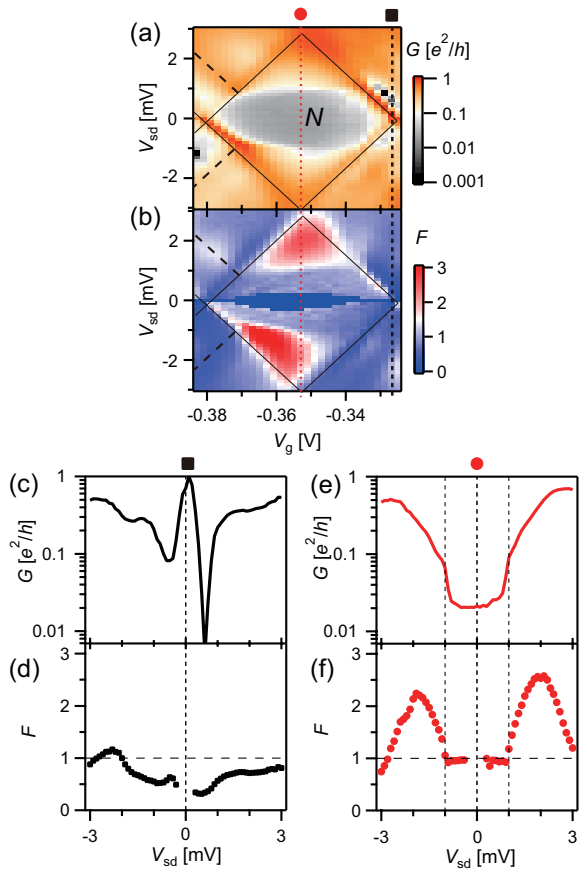}
 \end{center}
 \caption[Shot noise and Fano factor spectroscopy]{(Color online). (a) and (b): Color plots of (a) differential conductance $G$ and (b) Fano factor $F=\rss{S}{I}/2eI$ for the $N$-electron diamond as a function of $\rss{V}{sd}$ and $\rss{V}{g}$. We set $F=0$ in the region close to $\rss{V}{sd}=0$, where $I\sim 0$ and $\rss{S}{I}$ is below the resolution of our noise measurement system. (c) and (d) [(e) and (f)]: Differential conductance $G$ and Fano factor $F$ traced along the black (red) vertical dotted line in (a) and (b) corresponding to the sequential tunneling (cotunneling) regime. In (d) and (f), Poissonian Fano factor $F=1$ is shown as horizontal dashed lines. The vertical dashed lines at $\rss{V}{sd}=\pm 1$ mV in (e) and (f) indicate the inelastic cotunneling thresholds.}
 \label{figCotunneling03}
\end{figure}

Measuring the differential conductance as a function of $\rss{V}{g}$ and $\rss{V}{sd}$ reveals additional structures inside the $N$-electron diamond [Fig.~\ref{figCotunneling01}(c)], which are associated with cotunneling processes.
The horizontal steps at $\abs{\rss{V}{sd}}\approx 1$ meV, above which the conductance is strongly enhanced, represent the onset of inelastic cotunneling.
From the inelastic cotunneling thresholds, the energy difference between the ground and first excited states of the $N$-electron diamond is deduced to be $\Delta E = 1$ meV.
Below the thresholds ($\abs{e\rss{V}{sd}}<\Delta E$), we observe a small but finite conductance $G\approx 0.02$ $e^2/h$, indicating the elastic contribution [see also Fig.~\ref{figCotunneling03}(e)].
As we show below, the large energy splitting $\Delta E$ in our QD is essential for the observation of shot noise associated with the elastic cotunneling process.
These dc transport measurements also allow us to deduce other device parameters: the on-site Coulomb interaction $U=2$ meV is deduced from the width (3 meV) of the $N$-electron diamond (which equals $U+\Delta E$), and the dot-lead coupling strength $\Gamma=0.4$ meV from the width of the Coulomb peak.
We estimated the number of electrons in the QD to be $N\sim 10$ from the effective diameter of the QD ($\sim 40$ nm).

\begin{figure}
 \begin{center}
\includegraphics[scale=1]{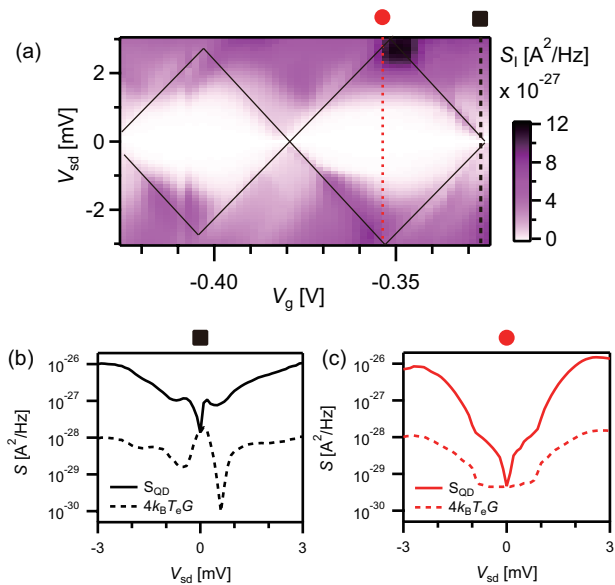}
 \end{center}
 \caption[Setup of inelastic cotunneling and DC spectroscopy]{(Color online). (a) Color plot of the shot noise $\rss{S}{I}$ as a function of $\rss{V}{sd}$ and $\rss{V}{g}$.  (b) [(c)] Total current noise $\rss{S}{QD}(\rss{V}{sd})$ and the estimated thermal noise $4\rss{k}{B}\rss{T}{e}G(\rss{V}{sd})$ with the base electron temperature $\rss{T}{e}=90$ mK at $\rss{V}{g}=-0.327$ V ($\rss{V}{g}=-0.353$ V) square (circle) in (a).}
 \label{figCotunneling02}
\end{figure}

%Measurement of the excess shot noise with respect to thermal noise
Shot noise $\rss{S}{I}(\rss{V}{sd})$ is obtained as a function of $\rss{V}{sd}$  by subtracting the thermal noise from the total current noise $\rss{S}{QD}(\rss{V}{sd})$ measured at each $\rss{V}{sd}$, i.e.,  $\rss{S}{I}(\rss{V}{sd}) = \rss{S}{QD}(\rss{V}{sd}) - 4\rss{k}{B}\rss{T}{e}G(\rss{V}{sd})$.
Here, the thermal noise is calculated from the electron temperature $\rss{T}{e}$ and the measured $G(\rss{V}{sd})$.
We estimate $\rss{T}{e}\approx 90$ mK via the relation $\rss{S}{QD}(0)=4\rss{k}{B}\rss{T}{e}G(0)$, which is expected to hold because $\rss{S}{I}(0)=0$.
The obtained $\rss{S}{I}$ is shown in Fig.~\ref{figCotunneling02}(a) for the same range of $\rss{V}{g}$ and $\rss{V}{sd}$ as in Figs.~\ref{figCotunneling01}(b) and (c).
We find that $\rss{S}{I}$ and $G$ [Fig.~\ref{figCotunneling01}(c)] show qualitatively similar behavior as a function of $\rss{V}{g}$ and $\rss{V}{sd}$.

In the above analysis, we assumed that $\rss{T}{e}$ was constant and independent of $\rss{V}{sd}$.
However, it is possible that for large $\abs{\rss{V}{sd}}$ the high current induces electron heating and enhances the thermal noise above the expected value, which would result in the overestimation of $\rss{S}{I}$.
To examine the validity of our analysis, in Figs.~\ref{figCotunneling02}(b) and \ref{figCotunneling02}(c), we compare the measured noise and the thermal noise estimated with $\rss{T}{e}=90$ mK (constant) for two representative cases: (b) the sequential tunneling regime ($\rss{V}{g}=-0.327$ V) and (c) the cotunneling regime ($\rss{V}{g}=-0.353$ V).
In both cases, the estimated thermal noise is about two orders of magnitude smaller than the measured noise in the relevant range, $\abs{\rss{V}{sd}}>1$ mV. 
This guarantees that the above analysis remains valid unless extremely strong heating leads to $\rss{T}{e}\gg 1$ K.
As we see below, the negligible contribution of the heating effect can also be confirmed by the Fano factor in the sequential tunneling regime.
Even with the current as high as $I>20$ nA at $\rss{V}{sd}=3$ mV, the measured Fano factor is within the expected range, $F<1$ [Fig.~\ref{figCotunneling03}(d)], indicating that the current-induced excess thermal noise is negligible compared with the shot noise.

Below, we mainly focus on the region of $N$-electron diamond.
Several intriguing features become evident when the Fano factor, calculated as $F(\rss{V}{g},\rss{V}{sd})=\rss{S}{I}(\rss{V}{g},\rss{V}{sd})/2eI(\rss{V}{g},\rss{V}{sd})$, is plotted as a function of $\rss{V}{g}$ and $\rss{V}{sd}$ [Fig.~\ref{figCotunneling03}(b)].
For comparison, the corresponding values of the differential conductance $G$ are shown in Fig.~\ref{figCotunneling03}(a).
In order to highlight the contrasting behavior of shot noise in different transport regimes, in Figs.~\ref{figCotunneling03}(c)-\ref{figCotunneling03}(f) we plot the measured conductance $G$ and Fano factor $F$ traced along the black and red vertical dotted lines in Figs.~\ref{figCotunneling03}(a) and (b).
First, a clear suppression of shot noise is observed in the sequential tunneling regime outside the Coulomb diamond, where the Fano factor varies between $0.5<F<1$ as a function of $\rss{V}{sd}$ [Fig.~\ref{figCotunneling03}(d)].
This sub-Poissonian noise for the sequential tunneling regime agrees with the reported value.\cite{aZhangPRL07, aKuznetsovPRB98}
Inside the diamond, we find Poissonian shot noise $F \approx 1$ (the measured value is $F= 0.95\pm0.05$) in the elastic cotunneling regime ($\abs{\rss{V}{sd}}<1$ mV) and super-Poissonian shot noise with the Fano factor $1<F<3$ in the inelastic cotunneling regime ($\abs{\rss{V}{sd}}>1$ mV) [Fig.~\ref{figCotunneling03}(f)].
It is noteworthy that the shot noise property can clearly distinguish the different transport dynamics in the sequential tunneling and inelastic cotunneling regimes, which is not apparent from the values of $G$ [Figs.~\ref{figCotunneling03}(c) and (e)].

\begin{figure}[b]
 \begin{center}
\includegraphics[scale=1]{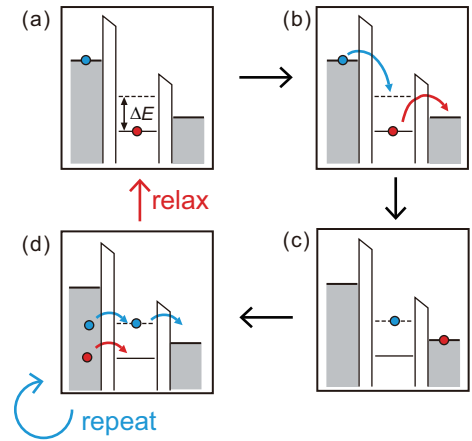}
 \end{center}
 \caption[Schematic of noise enhancement]{(Color online). Schematic illustrations of inelastic cotunneling and subsequent sequential tunneling events that lead to super-Poissonian shot noise.}
 \label{figCotunneling04}
\end{figure}

%In this paragraph, we describe the super-Poissonian noise in the inelastic cotunneling.
We now discuss in more detail the shot noise properties in the cotunneling regime [Fig.~\ref{figCotunneling03}(f)].
In the case of elastic cotunneling, the system returns to its ground state after each tunneling event.\cite{aThielmannPRB05, aWeymannPRB08}
This implies that only one electron is transported through each cycle.
Accordingly, $F=1$ is expected when elastic cotunneling is the dominant transport mechanism.
Indeed, our observation of Poissonian Fano factor $F\approx 1$ for $\abs{\rss{V}{sd}}<1$ mV agrees well with this expectation, demonstrating that the transport is dominated by elastic cotunneling in this regime.
In turn, the super-Poissonian Fano factor $1<F<3$ observed in the inelastic cotunneling regime ($\abs{\rss{V}{sd}}>1$ mV) implies that the transport occurs through a bunched flow of electrons involving on average $F$ electrons.
This Fano-factor enhancement can be understood by noting that inelastic cotunneling leaves the QD in the excited state, and so it must be followed by relaxation processes before the system returns to its original state [Figs.~\ref{figCotunneling04}(a)-(c)].
The possible energy relaxation processes include acoustic-phonon emission\cite{aFujisawaScience98} and sequential tunneling.\cite{aWeymannPRB08, aOnacPRL06, aAghassiAPL08} 
In the former, however, only one electron is emitted through each cycle.
In contrast, if inelastic cotunneling is followed by a subsequent sequential tunneling process, it happens that more than one electron is transported in each cycle before the system returns to the original state [Fig.~\ref{figCotunneling04}(d)].
Note that the sequential tunneling process through the excited state can occur repeatedly several times until the system eventually relaxes to the ground state (``repeat'' in the Fig.~\ref{figCotunneling04}).
The Fano factor is thus enhanced, reflecting the number of repetitions of this sequential tunneling process.
The observed value of $F\sim 2.5$ indicates that the subsequent sequential tunneling process is repeated on average 1.5 times, if relaxation via phonon emission is negligible.
For even larger $\rss{V}{sd}$ such that the transport window involves the single-particle levels in the QD, the Fano factor is expected to be suppressed to  the sub-Poissonian value, since the sequential tunneling becomes dominant.
Indeed, in Fig.~\ref{figCotunneling03}(f), the Fano factor is gradually suppressed to sub-Poissonian noise for $\abs{\rss{V}{sd}}>2$ mV.
%All results of the Fano facto qualitatively and quantitatively agrees well with the previous theoretical prediction.\cite{aWeymannPRB08}

% Here, we describe the discussion part

In conclusion, we have reported shot noise measurements on a semiconductor quantum dot in the cotunneling regime.
The large level spacing $\Delta E$ yielded cotunneling current high enough for the shot noise to be resolved in our measurement system.
We observed Poissonian shot noise $F\approx 1$ in the elastic cotunneling regime and super-Poissonian shot noise  $1<F<3$ in the inelastic cotunneling regime.
This result demonstrated that shot noise spectroscopy reveals the microscopic mechanisms involved in the cotunneling transport.

Our noise measurement system was constructed with reference to the setup built by K.~Kobayashi and M.~Hashisaka; we thank them for valuable discussions.
We thank I.~Mahboob for valuable comments on the manuscript.

\end{document}

%\preprint{}

\title{Supplementary Material for \\ ``Shot noise spectroscopy on a semiconductor quantum dot in the elastic and inelastic cotunneling regimes''}


\author{Yuma Okazaki}
\email[]{okazaki.y@lab.ntt.co.jp}
\affiliation{NTT Basic Research Laboratories, NTT Corporation, 3-1 Morinosato-Wakamiya, Atsugi, Kanagawa 243-0198, Japan}
\affiliation{Department of Physics, Tohoku University, Sendai, Miyagi 980-8578, Japan}
\author{Satoshi Sasaki}
\affiliation{NTT Basic Research Laboratories, NTT Corporation, 3-1 Morinosato-Wakamiya, Atsugi, Kanagawa 243-0198, Japan}
\affiliation{Department of Physics, Tohoku University, Sendai, Miyagi 980-8578, Japan}
\author{Koji Muraki}
\affiliation{NTT Basic Research Laboratories, NTT Corporation, 3-1 Morinosato-Wakamiya, Atsugi, Kanagawa 243-0198, Japan}

%\email[]{}
%\thanks{}


\date{\today}

%\begin{abstract}
%This supplementary material describes the details of the current noise measurements used in 
%\end{abstract}

%\keywords{}

\maketitle

This supplementary material describes the details of the current noise measurements presented in the main article ``Shot noise spectroscopy on a semiconductor quantum dot in the elastic and inelastic cotunneling regimes.''
To measure current noise sourced from the quantum dot (QD), we employ cryogenic-amplification and fast-Fourier-transform based current noise measurement system.\cite{ade-PicciottoNature98,  aDicarloRevSciInstl06,  aHashisakaJPhysConfSer08}
This system is calibrated with reference to thermal noise, and the estimated resolution is $\delta S \sim 2\times 10^{{-}29}$\,$\mathrm{A^2/Hz}$.

\begin{figure}
 \begin{center}
\includegraphics[scale=0.9]{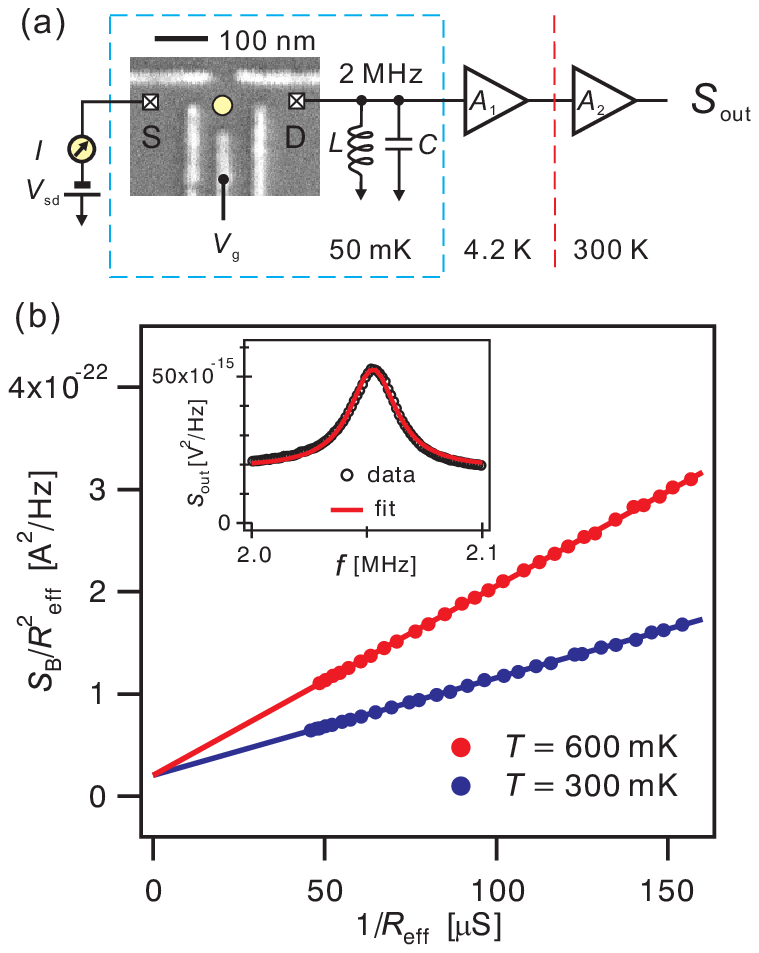}
 \end{center}
 \caption[Setup of inelastic cotunneling and DC spectroscopy]{ (a) Scanning electron microscope image of the device and schematic of the setup for measuring the current noise. (b) A plot of $\rss{S}{B}/\rss{R}{eff}^2$ versus $1/\rss{R}{eff}$ at temperatures $T=300$ and $600$\,mK. The solid lines are linear fits to the data. (Inset) Measured voltage power spectral density $\rss{S}{out}$ (circle) versus frequency $f$ and the Lorentzian fit to the data using Eq.~\ref{eqLorenz} (solid line). }
 \label{figCotunnelingS01}
\end{figure}

Figure \ref{figCotunnelingS01}(a) shows a scanning electron microscope image of our device and the setup for measuring the current noise.
The drain contact of the QD device is connected with the LC tank to form an RLC tank circuit, with `R' being the resistance of the QD.
The inductance $L\sim 15.5$\,$\mu\mathrm{H}$ and the capacitance $C\sim 390$\,pF provides the resonant frequency $ f_0=1/2\pi\sqrt{LC}\sim 2.05$\,MHz, at which the contribution of 1/f noise is negligible.
Here, the capacitance $C\sim 390$\,pF comes from a combination of a parasitic capacitance $\rss{C}{coax}\sim 80$\,pF of a coaxial cable connecting the device with the cryoamp and a ceramic chip capacitor $\rss{C}{chip}\sim 310$\,pF.
A voltage fluctuation across the RLC tank is amplified by a home-made cryoamp cooled to the liquid helium temperature (4.2\,K) with voltage power gain $A_1\sim 1$\,$\mathrm{V^2/V^2}$, followed by a room-temperature amplifier (NF SA-220F5) with voltage power gain $A_2\sim 4\times 10^{4}$\,$\mathrm{V^2/V^2}$.
A time-domain signal of the amplified voltage fluctuation is digitized and then is Fourier transformed to power spectral density.\cite{aDicarloRevSciInstl06}
The power spectral density is integrated over a period of 30 seconds.

In this system, the power spectral density $\rss{S}{out}(f)$ of the output voltage has a Lorentzian form, which can be expressed as 
\begin{equation}
	S_\mathrm{out}(f) = S_\mathrm{A} + \frac{ S_\mathrm{B}}{1+(f^2-f_0^2)^2/(f\Delta f)^2}.\label{eqLorenz}
\end{equation}
The measured $\rss{S}{out}$ agrees well with this Lorentzian formula, as shown in the inset of Fig.~\ref{figCotunnelingS01}(b).
Here, the width of the Lorentzian, $\Delta f$ in Eq.~\ref{eqLorenz}, relates to the effective impedance $\rss{R}{eff}$ of the RLC tank through $\Delta f = 1/2\pi\rss{R}{eff}C$.
Note that the value of $\rss{R}{eff}$ is usually smaller than $R$ of the QD because of an energy loss of the LC tank circuit.\cite{aDicarloRevSciInstl06, aHashisakaJPhysConfSer08}
Additionally, the height of the Lorentzian, $\rss{S}{B}$ in Eq.~\ref{eqLorenz}, is related to current noise via $\rss{S}{B}=A\rss{R}{eff}^2(\rss{S}{QD}+\rss{S}{Amp})$, where $A=A_1\cdot A_2$ is the total gain and $\rss{S}{QD}$ ($\rss{S}{Amp}$) is a current noise sourced from the QD (cryoamp).\cite{aZarchinPRB08, aYamauchiPRL11}

We calibrate two unknown parameters $\rss{S}{Amp}$ and $A$ through a measurement of thermal noise.\cite{ade-PicciottoNature98}
When $I=0$, $\rss{S}{QD}$ dose not include shot noise induced by a current flow, but includes thermal noise induced by a Brownian motion of electrons.
In this case,  $\rss{S}{QD}$ can be determined by two controllable parameters: effective impedance $\rss{R}{eff}$ and system temperature $T$ via $\rss{S}{QD}=4\rss{k}{B}T/\rss{R}{eff}$ with $\rss{k}{B}$ being the Boltzmann constant.
We measure this thermal noise for various $\rss{R}{eff}$ values and 12 different temperatures from 200 to 700\,mK at 50\,mK intervals, as shown in Fig.~\ref{figCotunnelingS01}(b) for $T=300$ and $600$\,mK.
In this figure, measured $\rss{S}{B}/\rss{R}{eff}^2$ is plotted as a function of $1/\rss{R}{eff}$ together with the linear fitting results (solid lines).
The measured data shows linear dependence on $1/\rss{R}{eff}$ with the common y-intercept.
Since this dependence can be expressed as $\rss{S}{B}/\rss{R}{eff}^2=A\rss{S}{Amp}+A\cdot 4\rss{k}{B}T/\rss{R}{eff}$,  the gain $A$ can be determined from the gradient of the plot, and then $\rss{S}{Amp}$ is determined from the y-intercept.
The determined values are $\rss{S}{Amp}=3.64\times 10^{-28}$\,$\mathrm{A^2/Hz}$ and $A=5.60\times 10^{4}$\,$\mathrm{V^2/V^2}$.
The resolution $\delta S$ in $\rss{S}{QD}$ was estimated to be $\delta S \sim 2\times 10^{{-}29}$\,$\mathrm{A^2/Hz}$, corresponding to a Poissonian noise value of 60\,pA.
We additionally found that the thermal noise linearly depends on the system temperature $\rss{T}{sys}$ above 200\,mK, and gradually deviates below this temperature due to the poor cooling power of the refrigerator.
We estimate the lowest electron temperature to be $\rss{T}{e}=90$\,mK from the lowest value of the thermal noise.\cite{aHashisakaJPhysConfSer08}

%\bibliography{bibs}